\documentclass[letterpaper, 10 pt, conference]{ieeeconf}
\IEEEoverridecommandlockouts
\usepackage{graphics}
\usepackage{amsmath}
\usepackage{amssymb}

\usepackage{multirow}
\usepackage{graphicx}
\usepackage{color}
\usepackage[short]{optidef}
\usepackage{algorithm}
\usepackage[noend]{algpseudocode}
\algnewcommand\algorithmicinput{\textbf{Input:}}
\algnewcommand\AlgInput{\item[\algorithmicinput]}
\algnewcommand\algorithmicoutput{\textbf{Output:}}
\algnewcommand\AlgOutput{\item[\algorithmicoutput]}
\algnewcommand\algorithmicforeach{\textbf{for each}}
\algdef{S}[FOR]{ForEach}[1]{\algorithmicforeach\ #1\ \algorithmicdo}
\algrenewcommand\algorithmicindent{1em}
\usepackage{tcolorbox}
\usepackage{subcaption}
\usepackage{tikz}
\usepackage{booktabs}
\usepackage{siunitx}
\usepackage{adjustbox}
\usepackage{url}
\usepackage{hyperref}

\title{\LARGE \bf
Congestion Mitigation Path Planning\\for Large-Scale Multi-Agent Navigation in Dense Environments
}

\author{Takuro Kato$^{1}$, Keisuke Okumura$^{2,3}$, Yoko Sasaki$^{2}$, Naoya Yokomachi$^{1}$%
\thanks{This work has been published in IEEE Robotics and Automation Letters (RA-L), 2025. Licensed under a Creative Commons Attribution 4.0 License. Digital Object Identifier: \href{https://doi.org/10.1109/LRA.2025.3597871}{10.1109/LRA.2025.3597871}}
\thanks{$^{1}$TICO-AIST Cooperative Research Laboratory for Advanced Logistics, National Institute of Advanced Industrial Science and Technology (AIST), 1-1-1, Umezono, Tsukuba, Japan %
        {\tt\footnotesize \{takuro.katou, naoya.yokomachi\}@aist.go.jp}}%
\thanks{$^{2}$Artificial Intelligence Research Center, National Institute of Advanced Industrial Science and Technology (AIST), 2-4-7 Aomi, Koto-ku, Tokyo, Japan %
        {\tt\footnotesize \{okumura.k, y-sasaki\}@aist.go.jp}}%
\thanks{$^{3}$Department of Computer Science and Technology, University of Cambridge, Cambridge, CB3 0FD, UK}%
}
\begin{document}

\maketitle
\thispagestyle{empty}
\pagestyle{empty}
\begin{abstract}
In high-density environments where numerous autonomous agents move simultaneously in a distributed manner, streamlining global flows to mitigate local congestion is crucial to maintain overall navigation efficiency.  
This paper introduces a novel path-planning problem, \textit{congestion mitigation path planning (CMPP)}, which embeds congestion directly into the cost function, defined by the usage of incoming edges along agents' paths.
CMPP assigns a flow-based multiplicative penalty to each vertex of a sparse graph, which grows steeply where frequently-traversed paths intersect, capturing the intuition that congestion intensifies where many agents enter the same area from different directions.
Minimizing the total cost yields a set of coarse-level, time-independent routes that autonomous agents can follow while applying their own local collision avoidance.  
We formulate the problem and develop two solvers: \emph{(i)} an exact \textit{mixed-integer nonlinear programming} solver for small instances, and \emph{(ii)} a scalable two-layer search algorithm, \textit{A-CMTS}, which quickly finds suboptimal solutions for large-scale instances and iteratively refines them toward the optimum.  
Empirical studies show that augmenting state-of-the-art collision-avoidance planners with CMPP significantly reduces local congestion and enhances system throughput in both discrete- and continuous-space scenarios.
These results indicate that CMPP improves the performance of multi-agent systems in real-world applications such as logistics and autonomous-vehicle operations.
\end{abstract}

\section{Introduction}
Modern industrial environments often involve scenarios where agents move autonomously toward their destinations.
For example, in logistics warehouses~\cite{Koster_2007}, human operators pick items in densely stocked conditions, adapting to inventory changes and order modifications.
In restaurants, human workers and mobile robots collaborate on tasks such as wayfinding and meal serving.
Other scenarios, including airport surface operations~\cite{Morris2016PlanningSA}, railway systems~\cite{chen2023flatland3}, and traffic control at autonomous intersections~\cite{Dresner2008AMA}, highlight the applicability of autonomous and distributed agent systems.

Such systems benefit from the increase in the number of agents as they concurrently process tasks.
However, without attention to coordination tactics, \emph{congestion} can degrade system performance as agents block each other's movement.
In the worst case, the entire system can get into a deadlock situation.
This poses the non-trivial challenge of mitigating congestion in an environment where agents operate in a (semi-)decentralized manner with minimal safety mechanisms such as collision avoidance.

\begin{figure}[t]
  \centering
  \includegraphics[width=\columnwidth]{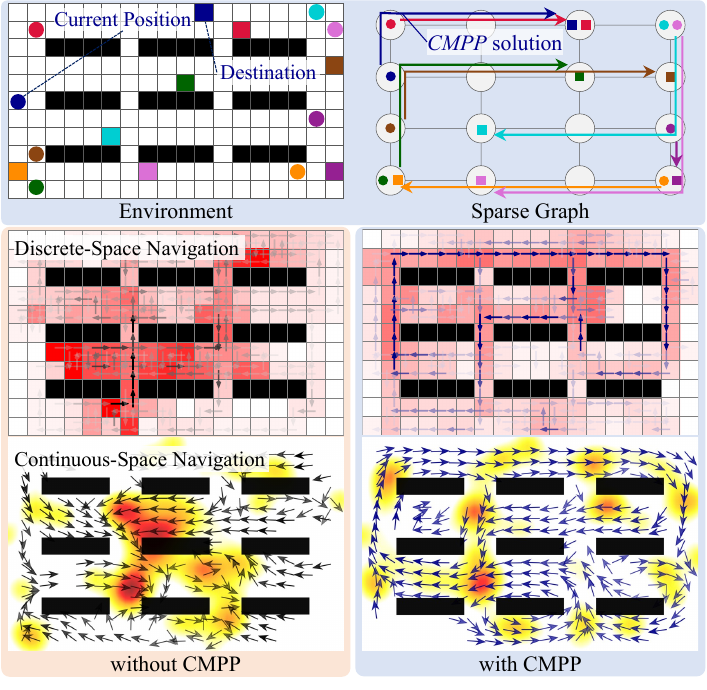}
  \caption{Overview of CMPP. \textbf{Top:} Environment with agents' current positions (circles) and destinations (squares), and the sparse graph used in CMPP. Multiple agents may share the same vertex in the graph. \textbf{Middle/Bottom:} Simulation snapshots in a discrete (middle) and a continuous (bottom) space with 60 agents; heatmaps indicate spatial occupancy, with higher intensity penalizing stuck agents. CMPP streamlines agent movements and mitigates congestion. Demonstration videos are provided as supplementary material and are accessible via IEEE Xplore.}
  \label{fig:overview}
\end{figure}    

Congestion mitigation has been explored across various domains. 
In freeway traffic management, strategies such as variable speed limits~\cite{Hegyi2005} have been proposed.
These methods target structured, lane-based networks, limiting their applicability to diverse environments.
Other work dynamically adjusts traversal costs between spatial regions~\cite{Chung2019}, but relies on pre-trained neural networks, which restrict adaptability. 
A different line of research focuses specifically on multi-agent path finding (MAPF)~\cite{Stern_SoCS2019}, optimizing congestion-aware guide paths at the grid level~\cite{Chen_AAAI2024}.

Inspired by these works, we formulate the \textit{congestion mitigation path planning (CMPP)} problem, which works effectively across diverse high-density navigation scenarios in both continuous and discrete spaces, without requiring pre-training.
We assume agents autonomously handle collision avoidance, making fine-grained path planning impractical.
Thus, we abstract the environment as a sparse graph, whose vertices represent key locations such as intersections or corridor entrances, to provide coarse-level route guidance, as shown in the top row of Fig.~\ref{fig:overview}.
On this graph, we propose a cost function defined by the usage of incoming edges along agent paths.
In our design, the cost at each vertex increases multiplicatively when frequently-traversed paths intersect, reflecting the intuition that congestion is severe in areas that many agents enter from different directions.
Minimizing the total cost yields a set of time-independent paths that naturally streamline agent flows and mitigate congestion, as demonstrated in the middle and bottom rows of Fig.~\ref{fig:overview}.

CMPP is related to MAPF, which assigns agents collision-free, time-dependent paths on a graph, typically minimizing total travel time.
While MAPF is a popular abstraction for warehouse automation~\cite{Wurman_2008}, real-world deployments introduce time inconsistencies due to communication delays, motion inaccuracies, and human intervention, making MAPF plans difficult to execute accurately.
Instead, CMPP abstracts path planning by identifying congestion-resistant routes suitable for online adjustment by autonomous agents.

Our contributions are \emph{(i)}~defining CMPP itself, \emph{(ii)}~proposing practical CMPP solvers, and \emph{(iii)}~demonstrating CMPP's applicability to discrete and continuous multi-agent navigation.
Specifically,
\emph{(i)}~we formulate CMPP as an optimization problem that minimizes overall congestion and promotes efficient agent flow.
This is based on quantifying the degree of congestion by measuring the overlap of agent paths.
\emph{(ii)}~We introduce two efficient solvers: 
an exact solver based on \textit{mixed-integer nonlinear programming (MINLP)} for small-scale instances, and a scalable two-layer search algorithm, \textit{anytime congestion mitigation tree search (A-CMTS)}, which quickly finds suboptimal solutions, refines them toward optimality.
As an application, \emph{(iii)} we demonstrate CMPP's practical effectiveness in both continuous and discrete domains:
integrating CMPP with the continuous-space collision-avoidance algorithm ORCA~\cite{vandenberg_springer11} raises the success rate of navigating 400 agents from 83.9 \% to 99.0 \%, while coupling CMPP with the discrete planner PIBT~\cite{okumura2022priority} in a lifelong-MAPF~\cite{Li_AAAI21lifelong} warehouse scenario yields up to a 58 \% throughput gain for 1,500 agents.


\section{Preliminaries}\label{sec:preliminaries}
\subsection{Assumptions}
\paragraph*{Map Abstraction}
We assume the environment is modeled as a sparse directed graph $G=(V,E)$, where each edge is bidirectional (if $(u, v) \in E$, then $(v, u) \in E$).
The sparse graph has vertices at key locations (e.g., intersections and corridors) instead of a fine-grained discretization of the entire space as shown in Fig.~\ref{fig:overview}.
Each vertex $v \in V$ represents a localized area, aggregating multiple nearby agents into a single node, simplifying the computation.

\paragraph*{Path}
Let $\pi := \{\pi_1,\ldots,\pi_n\}$ denote the set of agent paths on $G$, where each $\pi_i$ is an ordered sequence of vertices representing the path for agent $i$.
As CMPP does not explicitly control movement timing, paths are defined as sequences of vertices without specifying the arrival times.

\subsection{Problem Formulation} \label{sec:problem_formulation}

A \textit{CMPP instance} consists of a symmetric directed graph $G=(V,E)$, a set of agents $A=\{1,\ldots,n\}$,
and start-goal vertex pairs $\mathcal{S}=(s_1,\ldots,s_n)$ and $\mathcal{G}=(g_1,\ldots,g_n)$, where each $s_i$ and $g_i$ belongs to $V$.

Let $\pi_i=(\pi_i[1],\ldots,\pi_i[L_i])$ be the path of agent $i$, where $L_i$ is the path length.
A set of paths $\pi = \{\pi_1,\ldots,\pi_n\}$ is a \textit{solution} to the CMPP instance if each agent $i \in A$ satisfies the following constraints:
\begin{itemize}
\item \textit{Start and goal}: $\pi_i[1] = s_i \text{ and } \pi_i[L_i] = g_i$,
\item \textit{Adjacency}: Each step follows an edge, i.e., $(\pi_i[k], \pi_i[k+1]) \in E$ for all valid \textcolor{black}{$k \in\mathbb{N}^+$},
\item \textit{Simple path}: No vertex is revisited, i.e., $\pi_i[p] \neq \pi_i[q]$ for all $p\neq q$.
\end{itemize}

Our objective is to find a solution $\pi$ that minimizes \textit{total congestion cost} on $G$, as defined in the following subsection.

\subsection{Cost Function Design} \label{sec:cost_function}
\begin{figure}[t]
    \centering
    \includegraphics[width=0.8\linewidth]{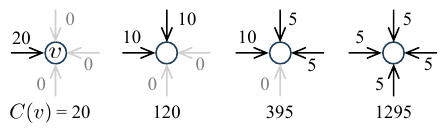}
    \caption{{\color{black}Example of changes in $C(v)$ when flow $f_e$ occurs on multiple edges in $\delta^-(v)$. The cost increases rapidly as the inflow routes become more dispersed.}}
    \label{fig:example}
\end{figure}

Congestion occurs when many agents enter the same vertex from different directions and cross or wait.
To quantify it, we first count the agents traversing each directed edge $e=(u,v)\in E$ by defining the \textit{flow} $f_e$ as:
\begin{align}
    f_e \!:=\! |\bigl\{i\in A \mid \exists k\in\mathbb{N}^+, (\pi_i[k],\pi_i[k+1])=(u,v)\bigr\}|.\label{eq:flow}
\end{align}
To penalize multi-directional merges, we define the \textit{congestion degree} $C(v)$ at a vertex $v$ as:
\begin{align}
    C(v) := \Bigl(\prod_{e\in\delta^{-}(v)} (f_e + 1)\Bigr) - 1, \label{eq:congestion}
\end{align}
where $\delta^-(v)$ is the set of directed edges entering $v$.
The multiplicative term in $C(v)$ increases sharply as flows arrive from multiple directions, while the $-1$ term ensures $C(v)=0$ when $v$ is unused.
As Fig.~\ref{fig:example} illustrates, 20 agents on one incoming edge yield $C(v)=20$, whereas distributing the same 20 agents over four edges raises $C(v)$ to $1295$.

CMPP aims to find a solution $\pi$ that minimizes the \textit{total congestion cost} $\sum_{v \in V} C(v)$.
Optimizing this objective encourages agents either to align in the same direction or to traverse well-separated areas, thereby suppressing cross-traffic bottlenecks.

\textbf{Note}:
Longer detours increase the number of traversed edges, thereby increasing the total congestion.
Therefore, reducing detours is implicitly encouraged.


\section{Related Work}\label{sec:related_work}
\paragraph*{Online Collision Avoidance Planners}
CMPP assumes that autonomous agents perform local collision avoidance.
Several online collision avoidance planners have been proposed in robotics, including PIBT~\cite{okumura2022priority} and collision shielding~\cite{Veerapaneni_ICAPS24} for discrete-space scenarios, as well as ORCA~\cite{vandenberg_springer11} and buffered Voronoi cells~\cite{Zhou_RAL17} for continuous-space settings.
These methods focus on local path adjustments and operate independently of higher-level global route planning.

\paragraph*{MAPF}
MAPF typically discretizes the environment into a fine grid and plans time-dependent, collision-free paths that minimize metrics such as total path length.
A well-known optimal solver is conflict based search (CBS)~\cite{Sharon_AAAI2012}, while numerous suboptimal solvers trade optimality for computational speed and scalability~\cite{li2021eecbs,okumura2023lacam}.
Variants such as lifelong-MAPF~\cite{Li_AAAI21lifelong} assign new goals to agents upon reaching their current destinations.
Unlike MAPF, CMPP shifts focus from strict collision avoidance and path-length minimization to streamlining agent flow and mitigating congestion by planning time-independent paths on a sparse graph.

\paragraph*{Traffic Management and Guide Heuristics}
Various approaches have been explored for congestion mitigation.
For example, in freeway traffic management, variable speed limits employing model predictive control have been studied~\cite{Hegyi2005}.
These methods assume structured, lane-based networks and do not generalize well to environments where agents move with greater freedom, such as warehouse situations.
For autonomous robot systems, methods that use pre-trained neural networks to dynamically adjust traversal costs between spatial regions have been proposed to guide agents toward congestion-avoiding routes~\cite{Chung2019}.
However, reliance on pre-training restricts adaptability to changes in graph topology or agent scale.
Several MAPF extensions have also been studied, including priority paths (highways)~\cite{Cohen_IJICAI16}, offline-optimized guidance graphs~\cite{Zhang_IJCAI2024}, uniform space utilization~\cite{Han_ICRA2022}, and delay predictions based on fleet histories~\cite{Yu_ICRA2023}.
Another notable method integrates \textit{traffic flow optimization (TFO)} to compute congestion-avoiding guide paths for MAPF planners~\cite{Chen_AAAI2024}.
These approaches remain specialized for discrete-space MAPF scenarios.

Unlike methods that rely on pre-training or offline-optimization, CMPP employs a sparse graph formulation that enables efficient online computation.  
Moreover, CMPP is compatible with both discrete-space and continuous-space online collision avoidance planners, providing flexibility across diverse navigation scenarios.

\section{Mathematical Programming Approach}\label{sec:MINLP}
{
\renewcommand{\arraystretch}{0.85}
\begin{table}[t]
\centering
\caption{
Stress test to find optimal solutions with MINLP, using \SI{60}{\second} timeout.
Each instance is prepared by randomizing the start and goal vertices.
``Timeout'' entries mean that no optimal solutions were found, while ``N/A'' means that no feasible solution was found.
}
\label{table:MINLP}
\begin{tabular}{llrr}
\toprule
\multicolumn{2}{l}{Map}                & $|A|$  & Time (\SI{}{\second}) \\ \midrule
\multirow{3}{*}{
\textit{3$\times$3 Grid}
} & \multirow{3}{*}{
\begin{minipage}{15mm}
\centering
\scalebox{0.4}{\includegraphics{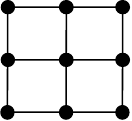}}
\end{minipage}
} & 10 & 1.2      \\
                  &                    & 20 & 12.3     \\
                  &                    & 30 & 39.6        \\ \midrule
\multirow{3}{*}{
\textit{10$\times$10 Grid}
} & \multirow{3}{*}{
\begin{minipage}{15mm}
\centering
\scalebox{0.35}{\includegraphics{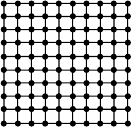}}
\end{minipage}
} & 100 & Timeout     \\
  &                    & 200 & Timeout       \\
  &                    & 300 & \textbf{N/A}        \\ \bottomrule
\end{tabular}
\end{table}
}
A straightforward approach to solving CMPP is to formulate it as a \textit{mixed-integer nonlinear programming (MINLP)} problem.
In this formulation, we define the following decision and auxiliary variables:
\begin{itemize}
\item $\displaystyle z_{i,e} \in \{0,1\}$ (decision): $z_{i,e} = 1$ if agent $i \in A$ traverses the directed edge $e \in E$, and $0$ otherwise.
\item $\displaystyle f_e \in {\color{black}\mathbb{Z}^{0+}}$ (auxiliary): flow on edge $e$.
\item $\displaystyle C(v) \in {\color{black}\mathbb{Z}^{0+}}$ (auxiliary): congestion degree at vertex $v$. 
\end{itemize}

The objective and constraints are formulated as follows:
\begin{mini!}[s]{}{\hspace{-2.5mm} \sum_{v \in V} C(v)}{}{\notag}
\addConstraint{}{\hspace{-2.5mm}f_e = \sum_{i\in A} z_{i,e}}
    {\hspace{2mm} \forall e \in E} \label{eq:cons_flow}
\addConstraint{}{\hspace{-2.5mm}C(v) \!= \!\prod_{e\in\delta^-(v)} \!(f_e \!+ \!1) \!- \!1}
    {\hspace{2mm} \forall v \in V} \label{eq:cons_congestion}    
\addConstraint{}{\hspace{-2.5mm}\sum_{e \in \delta^+(s_i) (\delta^-(g_i))} \!z_{i,e} = 1}
    {\hspace{2mm} \forall i \in A} \label{eq:cons_start_and_goal}
\addConstraint{}{\hspace{-2.5mm}\sum_{e \in \delta^-(v)} z_{i,e} = \sum_{e \in \delta^+(v)} z_{i,e}}
    {\hspace{2mm} \forall v \in V \hspace{-1mm} \setminus \hspace{-1mm}\{s_i, g_i\}, \forall i \in A} \label{eq:cons_connection}
\addConstraint{}{\hspace{-2.5mm}z_{i,(u,v)} + z_{i,(v,u)} \leq 1}
    {\hspace{2mm} \forall i \in A, \forall (u,v) \in E}. \label{eq:cons_redundant}  
\end{mini!}

Constraints~\eqref{eq:cons_flow} and \eqref{eq:cons_congestion} implement Eq.~\eqref{eq:flow} and Eq.~\eqref{eq:congestion}, respectively.
Once the binary variables $z_{i,e}$ are determined, the values of $f_e$ and $C(v)$ are uniquely defined by these constraints.
Constraint~\eqref{eq:cons_start_and_goal} enforces the start and goal conditions of CMPP, where $\delta^+(v)$ denotes the set of outgoing edges from vertex $v$.
Constraint~\eqref{eq:cons_connection} ensures adherence to the adjacency constraint, while constraint~\eqref{eq:cons_redundant} guarantees compliance with the simple path constraint.

\paragraph*{Discussion}
Although this approach efficiently finds exact solutions for small-scale instances, its computational cost becomes prohibitively high for larger problems.
Table~\ref{table:MINLP} summarizes the result of an experiment using SCIP~\cite{achterberg2009scip}, a popular mathematical solver for this optimization class.
On the \textit{$3\times3$ Grid}, the solver finds an optimal solution, but the runtime sharply increases as the number of agents grows.
On the \textit{$10\times10$ Grid}, the solver fails to compute an exact solution within the runtime limit for instances with $|A|=100$ or $|A|=200$ agents, although it still returns a feasible solution.
For $|A|=300$, however, it fails to return even a feasible solution within the same time limit.
These results indicate that the MINLP-based method has scalability limitations.
In the next section, we propose a search-based approach to overcome this limitation.

\begin{algorithm}[t]
\small
\caption{High-level Search of A-CMTS} \label{alg:high-level}
\begin{algorithmic}[1]
\AlgInput a CMPP instance $\{G, A, \mathcal{S}, \mathcal{G}\}$, suboptimal factor $\omega \in \mathbb{R}_{\ge 1}$
\AlgOutput a solution $\pi_{\text{Best}}$
\Procedure{High-level}{}
    \State $R \gets $ \textit{NewNode()} \Comment{$R$: root node}
    \State $R.C^+$ $\gets \emptyset$, $R.C^-$ $\gets \emptyset$  \Comment{Forced / forbidden constraints}
    \State $R$.$\pi$ $\gets$ Prioritize Planning (PP) using \Call{Low-level}{}()
    \State $R.\mathrm{cost}$ $\gets$ total congestion cost in $R.\pi$
    \State $Open \gets$ \textit{PriorityQueue()} \Comment{Ordered by cost (lowest first)}
    \State $Open.\text{push}(R)$
    \State $\pi_{\text{Best}} \gets$ $R$.$\pi$, $UB \gets R.\mathrm{cost}$ \Comment{$UB$: cost upper bound}

    \While{$Open \neq \emptyset$ $\land \neg$ interrupt()}
        \State $N \gets Open.$\text{pop()} \Comment{Node with lowest cost}
        \If {$N.\mathrm{cost} < UB$}
            \State $\pi_{\text{Best}}$ $\gets$ $N.\pi$, $UB \gets N.\mathrm{cost}$
        \EndIf         
        \State $N.\mathrm{LB}$ $\gets$ estimate lower bound of $N$ satisfying Eq.~\eqref{eq:LB}
        \If {$UB \le \omega \cdot N.\mathrm{LB}$}
            \State \textbf{continue} \Comment {Branch cut}
        \EndIf
        \State $v^* \gets \underset{v \in \Gamma(N)}{\operatorname{argmax}} \hspace{1mm} C(v)$ \Comment $\Gamma(N)$: defined as Eq.~\eqref{eq:Gamma}
        \If {$v^* = \text{Null}$}
            \State \textbf{continue} \Comment{No suitable vertex remains}
        \EndIf

        \State $a^* \gets$ select an agent that visits $v^*$ satisfying Eq.~\eqref{eq:agent_select}
        \State $e^* \gets$ the edge to $v^*$ used by agent $a^*$ 
        \State $(P, Q)$ $\gets$ \Call{ExpandNode}{$N, a^*, e^*, v^*$} \Comment{Algorithm ~\ref{alg:expand-node}}
        \State $Open.\text{push}(P),\; Open.\text{push}(Q)$
    \EndWhile
    \State \Return $\pi_{\text{Best}}$
\EndProcedure
\end{algorithmic}
\normalsize
\end{algorithm}
\begin{algorithm}[t]
\small
\caption{High-level Node Expansion of A-CMTS} \label{alg:expand-node}
\begin{algorithmic}[1]
\AlgInput node $N$, agent $a$, edge $e$, vertex $v$
\AlgOutput a pair of child nodes $(P, Q)$
\Procedure{ExpandNode}{}
    \State $P$ $\gets$ \text{duplicate} $N$ \Comment{Child with forced constraint}
    \State $P.C^+ \gets P.C^+ \cup \{(a, e)\}$
    \State $Q$ $\gets$ \text{duplicate} $N$ \Comment{Child with forbidden constraint}
    \State $Q.C^- \gets Q.C^- \cup \{(a, e)\}$
    \State $Q.\pi_a \gets$ replan the path of $a$ using \Call{Low-level}{}()
    \State \texttt{/* Local search for agents sharing $v$ */}
    \For{each agent $a'$ such that $a' \neq a$ and $v \in Q.\pi_{a'}$}
        \State $Q.\pi_{a'} \gets$ replan the path of $a'$ using \Call{Low-level}{}()
    \EndFor
    \State $Q$.cost $\gets$ total congestion cost in $Q.\pi$
    \State \Return $(P, Q)$
\EndProcedure
\end{algorithmic}
\normalsize
\end{algorithm}
\section{Anytime Congestion Mitigation Tree Search} \label{sec:anytime_alg}

We present \textit{anytime congestion mitigation tree search (A-CMTS)}, which employs a two-layer structure similar to CBS~\cite{Sharon_AAAI2012}.
A-CMTS is designed to efficiently find suboptimal solutions even for large problem instances.
At the high level, the algorithm identifies vertices with high congestion and the agents visiting them, and expands search nodes with different constraints (forced or forbidden) to partition the solution space.
At the low level, each agent's path is replanned to satisfy all constraints imposed by the high level.

\subsection{High-Level Search}
Each search node in A-CMTS contains: \emph{(i)} forced edges $C^+$, \emph{(ii)} forbidden edges $C^-$, \emph{(iii)} a set of paths $\pi$, \emph{(iv)} total congestion cost, and \emph{(v)} a lower bound $\mathrm{LB}$ on the solution.
Algorithm~\ref{alg:high-level} outlines the high-level procedure of A-CMTS.

The root node $R$ starts with empty constraints, and initial paths $R.\pi$ are computed by prioritized planning (PP)~\cite{Silver_2005} via the low-level search described in Sec.~\ref{sec:low-level} (lines 2--4).
The solution cost for $R.\pi$ is computed using Eq.~\eqref{eq:congestion} (line 5).
Node $R$ is pushed into priority queue $Open$, ordered by ascending cost (lines 6--7).
Best-known solution $\pi_{\text{Best}}$ and upper bound $UB$ are initialized using $R$ (line 8).
Throughout the search, $UB$ is updated whenever a lower-cost solution is found.

In the main loop (lines 9--22), A-CMTS employs branch-and-bound strategy to find the optimal solution.
First, the node $N$ with the lowest cost is extracted from $Open$ (line 10).
If the cost of $N$ is lower than current upper bound $UB$, $\pi_{\text{Best}}$ and $UB$ are updated (lines 11--12).
Next, a lower bound $N.\mathrm{LB}$ is computed (line 13).
Specifically, $N.\mathrm{LB}$ satisfies:
\begin{equation} \label{eq:LB}
    \mathrm{LB} \le \min \bigl\{\sum_{v\in \pi} C(v) \mid \pi \text{ satisfies } N.C^+, N.C^- \bigr\},
\end{equation}
meaning that $N.\mathrm{LB}$ guarantees a lower bound on the total congestion cost for any feasible solution derived from node $N$.
Node $N$ is pruned (lines 14--15) if $UB \le \omega \cdot N.\mathrm{LB}$, where $\omega \in \mathbb{R}_{\ge 1}$ controls the accuracy-speed trade-off: $\omega=1.0$ ensures optimality, while $\omega>1.0$ permits bounded suboptimality (see Sec.~\ref{sec:property} for details).

If node $N$ is not pruned, A-CMTS expands it.
First, the vertex $v^*$ with the highest congestion is selected from the set $\Gamma(N)$, which is defined as:
\begin{multline} \label{eq:Gamma}
\Gamma(N) := \bigl\{ v \mid \exists a \in A, \exists e \in (\delta^-(v) \cap N.\pi_a) \\ \text{such that } (a,e)\notin N.C^+ \bigr\}.
\end{multline}
In other words, $\Gamma(N)$ contains vertices where at least one agent's path can still be modified (line 16).
If no suitable vertex remains ($v^* = \text{Null}$), node $N$ is not expanded (lines 17--18).
Otherwise, among the agents whose paths pass through $v^*$, an agent $a^*$ is selected according to:
\begin{equation}\label{eq:agent_select}
a^* \in \bigl\{ a \in A \mid \exists e^* \in \delta^-(v^*) \cap N.\pi_a, (a,e^*) \notin N.C^+ \bigr\},
\end{equation}
as the one whose rerouting is expected to effectively reduce the total congestion cost (lines 19--20).

Algorithm~\ref{alg:expand-node} described in the next subsection imposes new constraints on agent $a^*$ and generates two child nodes, which are pushed into $Open$ (lines 21--22).
The search continues until $Open$ is empty or a stopping criterion (e.g., runtime limit) is met (line 9).

\textbf{Note}:
In this study, the priorities for PP at the root node (line 5) are assigned based on agent indices.
The lower bound $N.\mathrm{LB}$ (line 13) is estimated as:
\[
N.\mathrm{LB} = \text{Length}(\pi_{\text{min}}) + \Delta_C,
\]
where $\pi_{\text{min}}$ is the shortest feasible path satisfying all constraints in $N.C^+$ and $N.C^-$, and $\Delta_C$ represents the additional congestion cost induced by the forced edges in $N.C^+$, as these edges cannot be modified in subsequent replanning.
Improving the accuracy of this estimate while maintaining computational efficiency remains an open research direction.

\subsection{High-Level Node Expansion}

Algorithm~\ref{alg:expand-node} expands a node $N$ by adding constraints related to a specific edge $e$ used by agent $a$ to enter vertex $v$.
Two child nodes are created:
\emph{(i)} node $P$, where agent $a$ is forced to use edge $e$ (lines 2--3). \emph{(ii)} node $Q$, where agent $a$ is forbidden from using edge $e$ (lines 4--5).
These constraints partition the solution space into two distinct regions.

Child node $P$ inherits paths directly from node $N$, as forcing edge $e$ requires no immediate path replanning.
Forbidding edge $e$ in node $Q$ necessitates replanning agent $a$'s path (line 6).
Since replanning agent $a$ may affect congestion at vertex $v$, the paths of all other agents through $v$ are also replanned (lines 8--9).
This replanning does not affect the optimality guarantee, but the pilot studies observed a convergence speedup with tighter upper bounds.
Finally, the total congestion cost of node $Q$ is updated (line 10), and both child nodes $P$ and $Q$ are returned (line 11).

\subsection{Low-Level Search} \label{sec:low-level}
Given an agent $a$ and its associated constraints $C^+_a \subseteq C^+$ and $C^-_a \subseteq C^-$, the low-level planner returns a path $\pi_a$ that minimizes additional congestion at vertices while satisfying all constraints:
\begin{mini!}[s]{}{\sum_{(u, v) \in \pi_a} \Delta C(v)}
    {}{\notag}
\addConstraint{}{\pi_a[1] = s_a, \pi_a[L_a]=g_a}
    {} {\notag}
\addConstraint{}{(u,v) \!\notin\! C^-_a \ \forall (u, v) \!\in\! \pi_a, \; (u,v) \!\in\! \pi_a \ \forall (u, v) \!\in\! C^+_a} {\notag}
\end{mini!}
Here, $\Delta C(v)$ is the incremental congestion at vertex $v$ when edge $(u,v)$ is used by agent $a$, given the other agents' paths.

\textbf{Note}:
This optimization must determine the order in which forced edges are visited, making it analogous to the traveling salesperson problem~\cite{garey1979computers} and generally computationally intractable for real-time use.
Therefore, we adopt a simple approximation for efficiency:
\emph{(i)}~Sort the forced edges in $C^+_a$ by their Euclidean distance from the agent's current vertex $s_a$. 
\emph{(ii)}~Run Dijkstra's algorithm to compute the path from $s_a$ to the start of the first forced edge, minimizing $\sum \Delta C(v)$, then traverse that edge. Repeat this sequentially for each forced edge, computing the congestion-minimizing path from the end of the previously traversed edge.
\emph{(iii)}~Concatenate the resulting segments to obtain a solution $\pi_a$.

\subsection{Property}\label{sec:property}
A-CMTS has the following theoretical guarantee:

\textit{Theorem:}
For any suboptimality factor $\omega \ge 1.0$, A-CMTS returns a solution whose cost is at most $\omega \cdot c^*$, where $c^*$ is the optimal solution cost. When $\omega = 1.0$, A-CMTS returns an optimal solution.

\begin{proof}
For the sake of contradiction, assume that A-CMTS returns a solution with cost $c > \omega \cdot c^*$.
Let $N^*$ be a node whose constraints $N^*.C^+$ and $N^*.C^-$ do not exclude the optimal solution $\pi^*$.
By the definition of lower bound (see Eq.~\eqref{eq:LB}), we have $N^*.\mathrm{LB} \leq c^*$.

If $N^*$ was pruned, then the upper bound at the time of pruning $UB'$ satisfies:
\begin{equation*}
   c \le UB' \leq \omega \cdot N^*.\mathrm{LB}.
\end{equation*}
Combining these inequalities yields:
\begin{equation*}
  c \le \omega \cdot c^*,
\end{equation*}
which contradicts our assumption that $c > \omega \cdot c^*$.
On the other hand, if $N^*$ was not pruned, A-CMTS continues to partition the solution space and eventually explores $\pi^*$.
Thus, the cost of the returned solution $c$ is always at most $\omega \cdot c^*$.

When $\omega=1.0$, we have $c \le c^*$.
Since no solution can have a cost lower than $c^*$, it follows that $c=c^*$.
\end{proof}


\subsection{Lifelong Variant}
In applications such as lifelong-MAPF or \textit{multi-agent pickup and delivery}~\cite{Ma_AAMAS2017}, where agents are continuously assigned new destinations upon reaching their current goals, A-CMTS must repeatedly solve CMPP instances.
To facilitate this, we initialize the root node of each A-CMTS computation with the solution paths computed in the previous calculation, denoted by $\pi_{\text{Prev}}$.
In other words, $\pi_{\text{Prev}}$ is reused as the solution path of the root node, allowing the algorithm to converge to stable solutions efficiently with minimal disruption between iterations.


\section{Experiments} \label{sec:experiments}
{
\tabcolsep = 4.2pt
\renewcommand{\arraystretch}{0.85}
\begin{table*}[t]
\centering
\caption{Performance of MINLP and A-CMTS for solving CMPP. \textit{Success} denotes the percentage of 50 runs yielding feasible solutions within \SI{1}{\minute}. \textit{Obj.} and \textit{Time} indicate the average objective values and computation times (\si{\second}), respectively. For A-CMTS, \textit{Initial solution} shows the values from the PP-based initial solution, while \textit{Final solution} gives the values at termination. \textit{Impr.} represents the average cost improvement (\%) from the initial to final solution.}
\label{table:optimization_experiment}
\begin{tabular}{@{}crrrrrrrrrrrrrr@{}}
\toprule
\multirow{4}{*}{Map} & \multicolumn{1}{c}{\multirow{4}{*}{$|V|$}} & \multicolumn{1}{c}{\multirow{4}{*}{$|A|$}} & \multicolumn{3}{c}{MINLP} & \multicolumn{9}{c}{A-CMTS} \\ \cmidrule(l){4-6} \cmidrule(l){7-15}
 & \multicolumn{1}{c}{} & \multicolumn{1}{c}{} & \multicolumn{1}{c}{\multirow{3}{*}{Success (\%)}} & \multicolumn{1}{c}{\multirow{3}{*}{Obj.}} & \multicolumn{1}{c}{\multirow{3}{*}{Time (\si{\second})}} & \multicolumn{1}{c}{\multirow{3}{*}{Success}} & \multicolumn{2}{c}{Initial solution} & \multicolumn{6}{c}{Final solution} \\ \cmidrule(l){8-9}\cmidrule(l){10-15}  
 & \multicolumn{1}{c}{} & \multicolumn{1}{c}{} & \multicolumn{1}{c}{} & \multicolumn{1}{c}{} & \multicolumn{1}{c}{} & \multicolumn{1}{c}{} & \multicolumn{1}{c}{\multirow{2}{*}{Obj.}} & \multicolumn{1}{c}{\multirow{2}{*}{Time}} & \multicolumn{3}{c}{$\omega=1.0$} & \multicolumn{3}{c}{$\omega=1.3$} \\ \cmidrule(l){10-12}\cmidrule(l){13-15} 
 & \multicolumn{1}{c}{} & \multicolumn{1}{c}{} & \multicolumn{1}{c}{} & \multicolumn{1}{c}{} & \multicolumn{1}{c}{} & \multicolumn{1}{c}{} & \multicolumn{1}{c}{} & \multicolumn{1}{c}{} & \multicolumn{1}{c}{Obj.} & \multicolumn{1}{c}{Impr. (\%)} & \multicolumn{1}{c}{Time} & \multicolumn{1}{c}{Obj.} & \multicolumn{1}{c}{Impr.} & \multicolumn{1}{c}{Time} \\ \midrule
\textit{3$\times$3 Grid} & \multirow{5}{*}{9} & 10 & \textbf{100} & \textbf{28.0} & 0.8 & \textbf{100} & 31.2 & \textbf{0.0} & \textbf{28.0} & \textbf{10.1} & 32.2 & 28.2 & 9.4 & 18.1 \\
\multirow{4}{*}{\begin{minipage}{15mm} \centering \scalebox{0.5}{\includegraphics{figure/3x3grid.pdf}} \end{minipage}} &  & 15 & \textbf{100} & \textbf{47.7} & 2.8 & \textbf{100} & 57.9 & \textbf{0.0} & 48.5 & 16.2 & 60.0 & 48.4 & \textbf{16.4} & 22.3 \\
 &  & 20 & \textbf{100} & \textbf{70.4} & 19.0 & \textbf{100} & 90.7 & \textbf{0.0} & 74.4 & 18.0 & 60.0 & 72.8 & \textbf{19.7} & 50.3 \\
 &  & 25 & \textbf{100} & \textbf{95.4} & 48.7 & \textbf{100} & 131.0 & \textbf{0.0} & 107.2 & 18.1 & 60.0 & 102.0 & \textbf{22.1} & 58.3 \\
 &  & 30 & \textbf{100} & \textbf{123.8} & 58.1 & \textbf{100} & 177.0 & \textbf{0.0} & 145.7 & 17.7 & 60.0 & 143.4 & \textbf{19.0} & 60.0 \\ \midrule
\textit{Connected} & \multirow{5}{*}{12} & 20 & \textbf{100} & \textbf{176.1} & 0.2 & \textbf{100} & 200.7 & \textbf{0.0} & 179.5 & \textbf{10.6} & 60.0 & 179.7 & 10.5 & 15.2 \\
\multirow{4}{*}{\begin{minipage}{15mm} \centering \scalebox{0.5}{\includegraphics{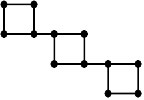}} \end{minipage}} &  & 40 & \textbf{100} & \textbf{569.7} & 0.4 & \textbf{100} & 694.9 & \textbf{0.0} & 596.7 & 14.1 & 60.0 & 590.8 & \textbf{15.0} & 45.0 \\
 &  & 60 & \textbf{100} & \textbf{1,191.7} & 0.6 & \textbf{100} & 1,556.4 & \textbf{0.0} & 1,294.8 & 16.8 & 60.0 & 1,273.1 & \textbf{18.2} & 60.0 \\
 &  & 80 & \textbf{100} & \textbf{2,018.7} & 0.8 & \textbf{100} & 2,728.7 & \textbf{0.0} & 2,248.6 & 17.6 & 60.0 & 2,212.7 & \textbf{18.9} & 60.0 \\
 &  & 100 & \textbf{100} & \textbf{3,048.6} & 1.1 & \textbf{100} & 4,217.0 & \textbf{0.0} & 3,590.2 & 14.9 & 60.0 & 3,561.8 & \textbf{15.5} & 60.0 \\ \midrule
\textit{10$\times$10 Grid} & \multirow{5}{*}{100} & 50 & \textbf{100} & 1,106.6 & 60.0 & \textbf{100} & 623.0 & \textbf{0.0} & \textbf{578.8} & \textbf{7.1} & 60.0 & \textbf{579.3} & 7.0 & 60.0 \\
\multirow{4}{*}{\begin{minipage}{15mm} \centering \scalebox{0.5}{\includegraphics{figure/10x10grid.pdf}} \end{minipage}} &  & 100 & 96 & 6,928.9 & 60.0 & \textbf{100} & 1,926.6 & \textbf{0.0} & \textbf{1,774.1} & \textbf{7.9} & 60.0 & 1,774.1 & \textbf{7.9} & 60.0 \\
 &  & 150 & 76 & 24,232.2 & 60.0 & \textbf{100} & 4,009.2 & \textbf{0.0} & \textbf{3,720.4} & \textbf{7.2} & 60.0 & \textbf{3,720.3} & \textbf{7.2} & 60.0 \\
 &  & 200 & 66 & 66,493.7 & 60.0 & \textbf{100} & 6,831.8 & \textbf{0.0} & \textbf{6,403.9} & \textbf{6.3} & 60.0 & \textbf{6,403.9} & \textbf{6.3} & 60.0 \\
 &  & 250 & 60 & 130,480.6 & 60.0 & \textbf{100} & 10,375.4 & \textbf{0.0} & \textbf{9,743.9} & \textbf{6.1} & 60.0 & \textbf{9,743.9} & \textbf{6.1} & 60.0 \\ \midrule
\textit{lak303d} & \multirow{5}{*}{265} & 100 & 32 & 64,068.4 & 60.0 & \textbf{100} & 13,394.2 & \textbf{0.0} & \textbf{12,500.4} & \textbf{6.7} & 60.0 & \textbf{12,500.4} & \textbf{6.7} & 60.0 \\
\multirow{4}{*}{\begin{minipage}{15mm} \centering \scalebox{0.13}{\includegraphics{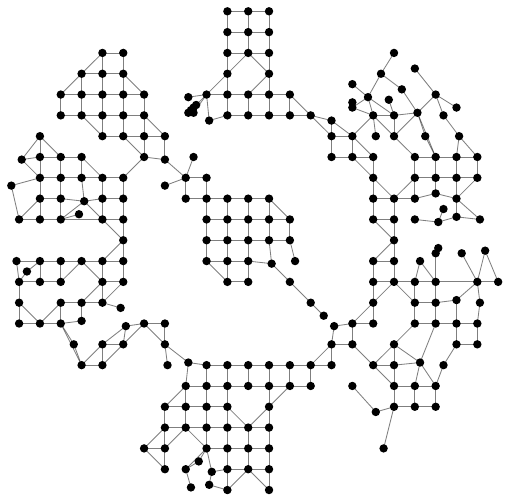}}\end{minipage}} &  & 200 & 2 & 529,492.0 & 60.0 & \textbf{100} & 55,518.8 & \textbf{0.0} & \textbf{50,992.8} & \textbf{8.2} & 60.0 & \textbf{50,992.8} & \textbf{8.2} & 60.0 \\
 &  & 300 & 0 & N/A & N/A & \textbf{100} & 135,029.9 & \textbf{0.1} & \textbf{122,897.7} & \textbf{9.0} & 60.0 & \textbf{122,897.7} & \textbf{9.0} & 60.0 \\
 &  & 400 & 0 & N/A & N/A & \textbf{100} & 260,542.0 & \textbf{0.1} & \textbf{236,333.3} & \textbf{9.3} & 60.0 & \textbf{236,333.3} & \textbf{9.3} & 60.0 \\
 &  & 500 & 0 & N/A & N/A & \textbf{100} & 450,215.5 & \textbf{0.1} & \textbf{400,370.1} & \textbf{11.1} & 60.0 & \textbf{400,370.1} & \textbf{11.1} & 60.0 \\ \bottomrule
\end{tabular}
\end{table*}
}
We evaluate CMPP through three sets of experiments.  
First, we compare two complementary solvers: \textit{MINLP}, an exact method for small instances, and \textit{A-CMTS}, a bounded-suboptimal algorithm for large instances.  
Second, we demonstrate CMPP's practical effectiveness in continuous-space scenarios by integrating A-CMTS with the online collision-avoidance planner \textit{ORCA}.  
Third, we validate our approach in discrete, lifelong-MAPF scenarios by coupling A-CMTS with the grid-based planner \textit{PIBT}.  
All experiments are conducted on a MacBook Pro (Apple M3 Max, 128 GB RAM).
MINLP computations are performed using SCIP, while A-CMTS is implemented in C++.
For ORCA and PIBT, we use publicly available implementations provided by the original authors.\footnote{We use the RVO2 library (\url{https://gamma.cs.unc.edu/RVO2/}) for ORCA, and the PIBT component from the author's LaCAM3 repository (\url{https://github.com/kei18/lacam3}).}

\subsection{Performance Evaluation of CMPP Solvers}
\paragraph*{Setup}
We evaluate the runtime, solution quality, and scalability of MINLP and A-CMTS under two suboptimality settings ($\omega=1.0$ and $\omega=1.3$).
Table~\ref{table:optimization_experiment} includes the sparse graph used for benchmarking.
For each test setting, 50 problem instances are generated by randomly assigning start and goal locations to the agents.
Each solver is given a \SI{1}{\minute} runtime limit per instance.

\begin{figure}[t]
  \centering
  \begin{subfigure}{0.49\columnwidth}
    \centering
    \includegraphics[width=\linewidth]{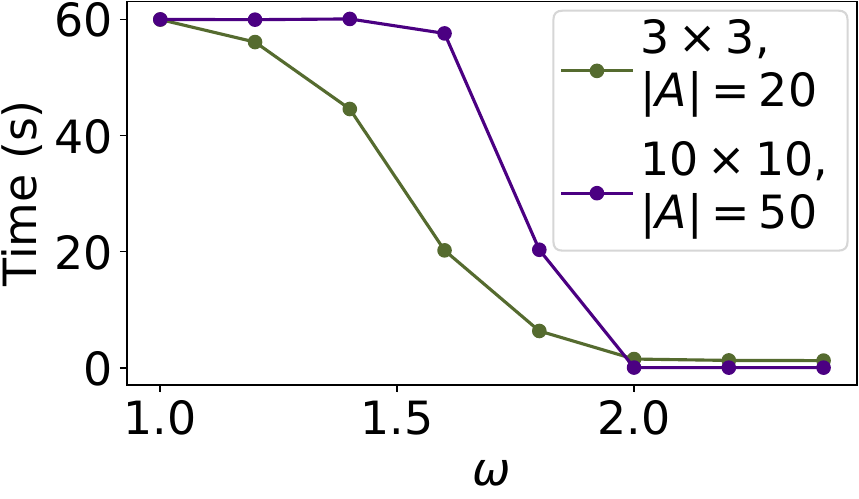}
  \end{subfigure}
  \begin{subfigure}{0.49\columnwidth}
    \centering
    \includegraphics[width=\linewidth]{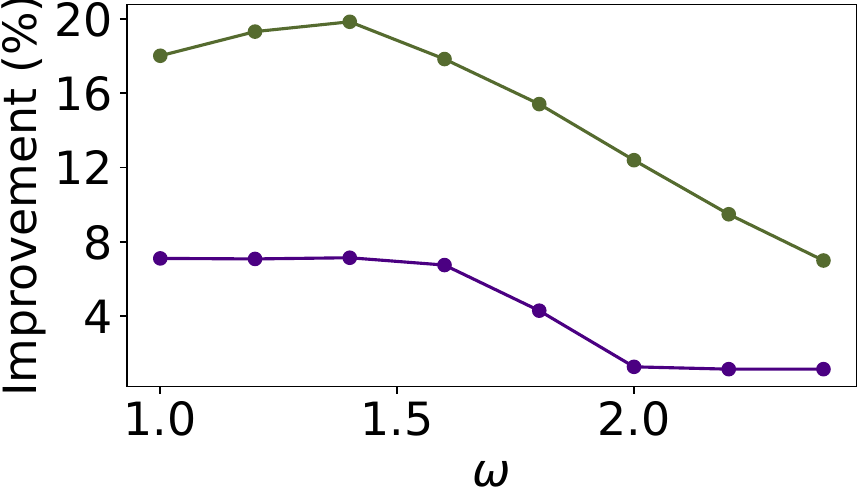}
  \end{subfigure}
    \caption{Effect of suboptimality factor $\omega$ on runtime (\si{\second}) and improvement ($\%$) from initial to final solution of A-CMTS.}
  \label{fig:suboptimal_ACMTS}
\end{figure}
{
\tabcolsep = 4.2pt
\renewcommand{\arraystretch}{0.85}
\begin{table}[t]
\centering
\caption{Performance of A-CMTS ($\omega=1.0$) for large-scale instances. \textit{Init.} indicates the PP-based initial solution.}
\label{table:scalability_ACMTS}
\begin{tabular}{@{}crrrr@{}}
\toprule
\multirow{2}{*}{Grid} & \multicolumn{1}{c}{\multirow{2}{*}{$|A|$}} & \multicolumn{1}{c}{Init.} & \multicolumn{2}{c}{Final} \\ \cmidrule(l){3-3} \cmidrule(l){4-5}
 & \multicolumn{1}{c}{} & \multicolumn{1}{c}{Time {\color{black}(\si{\second})}} & \multicolumn{1}{c}{Improvement {\color{black}(\%)}} & \multicolumn{1}{c}{Explored nodes} \\ \midrule
\multirow{3}{*}{\begin{tabular}[c]{@{}c@{}}10$\times$10\\ ($|V|$=100)\end{tabular}} & 2,000 & 0.3 & 2.3 & 7,441.4 \\
 & 6,000 & 2.6 & 1.7 & 2,573.5 \\
 & 10,000 & 7.3 & 0.8 & 1,441.5 \\ \midrule
\multirow{3}{*}{\begin{tabular}[c]{@{}c@{}}50$\times$50\\ ($|V|$=2,500)\end{tabular}} & 2,000 & 4.8 & 0.9 & 954.5 \\
 & 6,000 & 19.4 & 0.2 & 355.8 \\
 & 10,000 & 37.5 & 0.1 & 210.9 \\ \bottomrule
\end{tabular}
\end{table}
}
\begin{figure}[t]
  \centering
  \renewcommand{\arraystretch}{0.85}
  \begin{tabular}{@{}p{0.42\linewidth} p{0.52\linewidth}@{}}
    \begin{minipage}[t]{\linewidth}
      \vspace{0pt}\centering
      \includegraphics[width=0.83\linewidth]{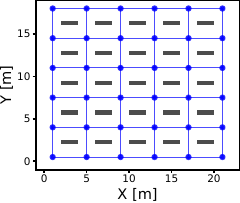}
    \end{minipage}
    &
    \begin{minipage}[t]{\linewidth}
      \centering\small%
      \setlength{\tabcolsep}{1pt}
      \begin{tabular}[t]{@{}l r@{}}
        \toprule
        Map size        & 22$\times$18.5 (m)\\
        Obstacle size   & 2$\times$0.5 (m)\\
        Agent radius    & 0.3 (m)\\
        Max velocity    & 2.0 (m/s)\\
        Simulation horizon & 60 (s)\\
        Time step       & 0.05 (s)\\
        \bottomrule
      \end{tabular}
    \end{minipage}
    \\
    \multicolumn{2}{c}{%
      \includegraphics[width=0.97\linewidth]{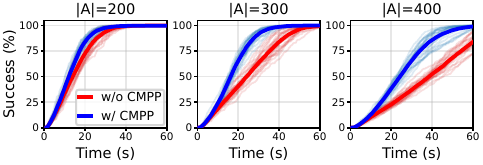}%
    }\\
  \end{tabular}
  \caption{{\color{black} Continuous-space guidance experiment. \textbf{Top:} Environment, sparse graph, and simulation parameters. \textbf{Bottom:} Success rate for $|A|=200, 300, 400$ with vanilla ORCA (red) and CMPP-guided ORCA (blue). Thin curves show individual results from 25 trials; bold curves are their mean. }}
  \label{fig:ORCA_experiment}
\end{figure}
{
    \setlength{\tabcolsep}{0pt}
    \newcommand{\entry}[5]{%
      \begin{minipage}{0.132\linewidth}
        \centering
        {\setlength{\baselineskip}{7pt}
         \scriptsize
         \begin{minipage}[t][\dimexpr4\baselineskip\relax][c]{0.80\linewidth}
           \raggedright
           \textit{#1}\\
           #2 (#3)\\
           #4 [#5\%]%
         \end{minipage}}\\[1pt]
      \IfFileExists{result_lifelong-mapf/map/#1.pdf}
      {\includegraphics[width=0.95\linewidth]{result_lifelong-mapf/map/#1.pdf}\vspace{0.3mm}}
      {\includegraphics[width=0.95\linewidth]{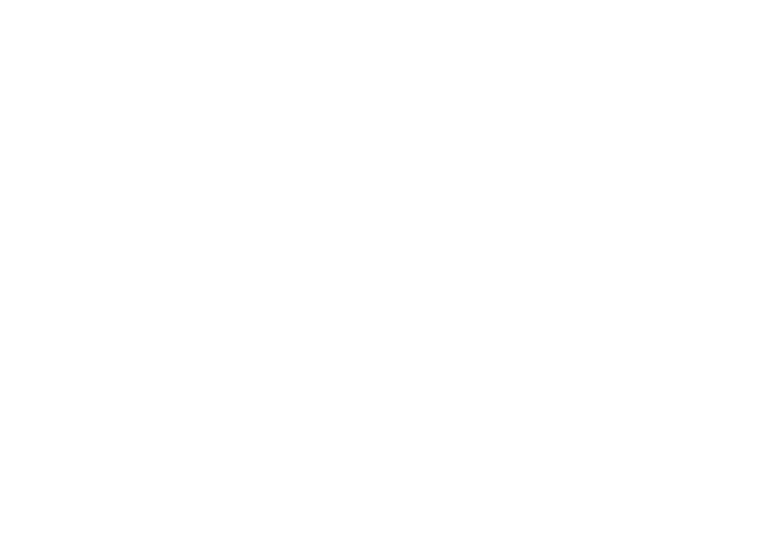}}
      \IfFileExists{result_lifelong-mapf/graph/#1.pdf}
      {\includegraphics[width=0.9\linewidth]{result_lifelong-mapf/graph/#1.pdf}}
      {\includegraphics[width=0.9\linewidth]{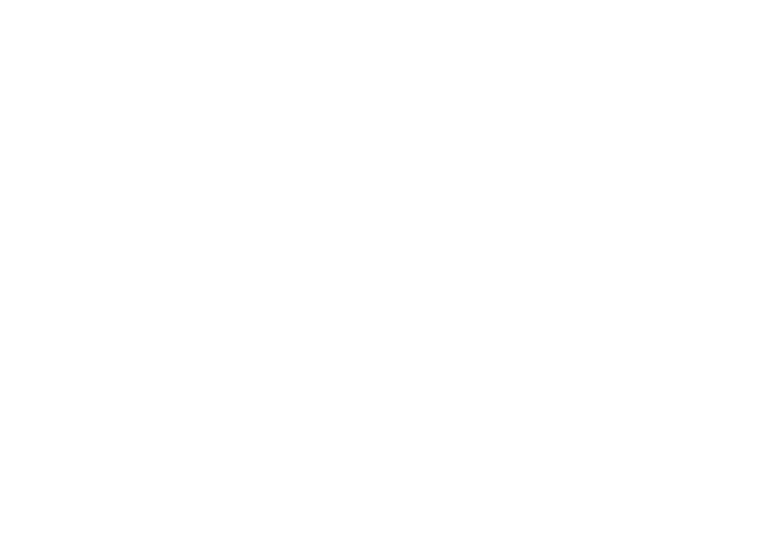}}
      \IfFileExists{result_lifelong-mapf/throughput/#1.pdf}
        {\includegraphics[width=\linewidth,trim=5pt 0pt 3pt 0pt,clip]{result_lifelong-mapf/throughput/#1.pdf}}
      {\includegraphics[width=\linewidth]{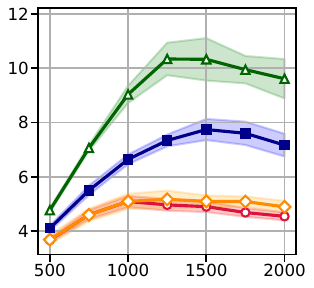}}      
    \end{minipage}
  }
  \newcommand{\labels}{
    \begin{minipage}{0.03\linewidth}
      \begin{tikzpicture}
        \node[rotate=90,font=\small](l1) at (0, 0.48) {Throughput};
        \node[rotate=90,font=\small](l2) at (0, 2.3) {Graph};
        \node[rotate=90,font=\small](l3) at (0, 4.0) {Map};
        \node[] at (0, -1) {};
        \node[] at (0, 6) {};
      \end{tikzpicture}
    \end{minipage}
  }
    \begin{figure*}[th!]
    \centering
    \begin{tabular}{cccccccc}
      \labels &
      \entry{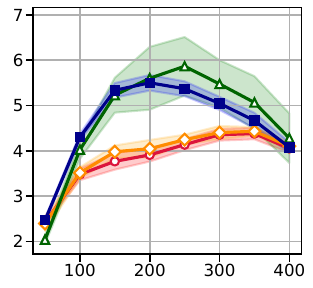}{$|V^*|$=422}{19x26}{$|V|$=130}{30.8} &
      \entry{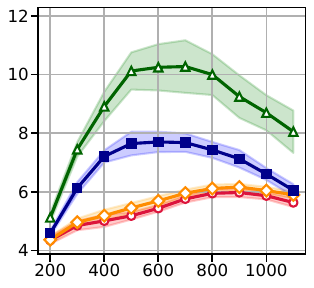}{1,278}{33x46}{128}{10.0} &
      \entry{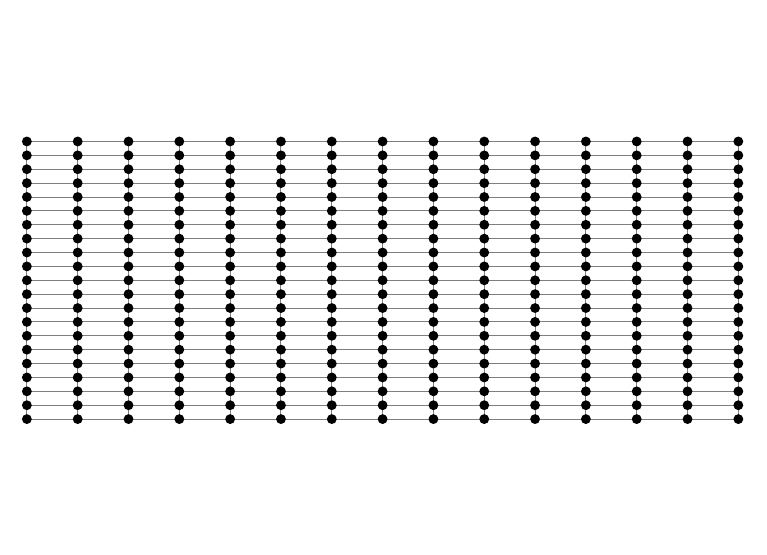}{5,699}{161x63}{315}{5.5} &
      \entry{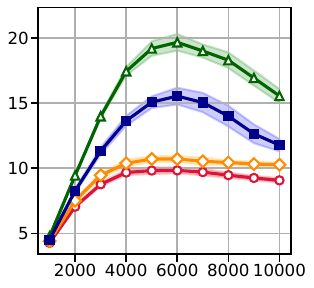}{22,599}{321x123}{1,189}{5.3} &
      \entry{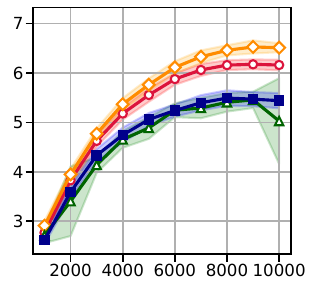}{14,784}{194x194}{265}{1.8} &
      \entry{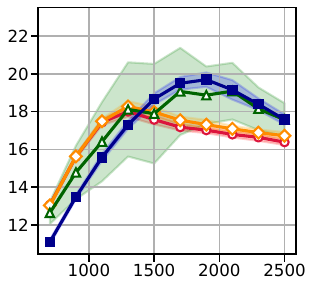}{3,687}{64x64}{463}{12.5} &
      \entry{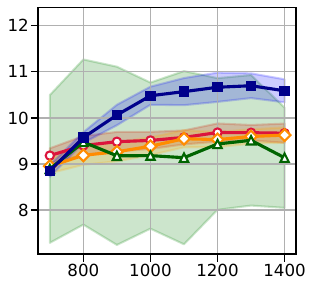}{3,270}{64x64}{476}{14.6}\\[-8pt]
      \multicolumn{8}{c}{\small Agents:~$|A|$}\\
      \multicolumn{8}{c}{\includegraphics[scale=0.9]{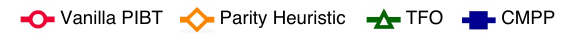}}
    \end{tabular}
    \caption{Results of applying CMPP to lifelong-MAPF setting. Numbers above each map indicate the map size $|V^*|$, the sparse graph size $|V|$, and the reduction ratio $|V|/|V^*|$ (\%). \textit{Graph} shows the graph used as input to CMPP, and \textit{Throughput} is the total number of arrivals normalized by the simulation duration ($500$ steps). Each plotted line represents the average of $25$ random scenarios for each setting, with the shaded regions indicating standard deviations.}
    \label{fig:applied_experiment_results}
  \end{figure*}
}

\paragraph*{Results}
Table~\ref{table:optimization_experiment} summarizes the results.
While MINLP efficiently finds optimal solutions on small-scale instances, its success rate declines for larger instances.
On larger maps, MINLP often returns suboptimal solutions or fails to find feasible solutions within the runtime limit.
In contrast, A-CMTS consistently achieves a $100\%$ success rate across all tests, benefiting from its two-level structure: a fast PP-based initial solution followed by iterative refinement.
On small-scale maps, the initial solutions computed by A-CMTS can be up to $43\%$ costlier than those from MINLP.
However, iterative refinements significantly reduce this gap to within $0-17.8\%$.
On larger-scale maps, A-CMTS not only produces superior initial solutions compared to MINLP within 0.1 \SI{}{\second} but also further improves them, ultimately obtaining solutions up to $11.1\%$ better upon termination.

\paragraph*{Impact of Suboptimality Factor}
Table~\ref{table:optimization_experiment} also shows that A-CMTS with $\omega=1.3$ sometimes terminates earlier than with $\omega=1.0$, while still providing comparable or even better solution quality.
A higher $\omega$ value enables more aggressive pruning in the high-level search, which enables faster progress through the search space within the limited time frame.
Figure~\ref{fig:suboptimal_ACMTS} shows the relationship between $\omega$, computation time, and solution quality.
These results highlight a flexible trade-off between solution optimality and computational efficiency, making A-CMTS practical for a variety of real-world applications.

\paragraph*{Scalability of A-CMTS}
Table~\ref{table:scalability_ACMTS} presents the scalability of A-CMTS for the number of agents $|A|$ and vertices $|V|$.
When $|V|=100$, A-CMTS rapidly computes initial solutions for instances with up to $10,000$ agents, achieving cost improvements ranging from $0.8\%$ to $2.3\%$ through iterative refinement.
Given the scale of the problems ($2,000$ to $10,000$ agents), improvements of a few percentage points represent significant performance gains for the overall system.
For $|V|=2,500$, although initial solution times increase with the number of agents, the refinement step still achieves consistent, albeit modest, cost improvements.
These results confirm that A-CMTS scales effectively to large-scale problems.

\subsection{Multi-Agent Guidance Simulation in Continuous Space} \label{sec:app_ORCA}
We evaluate CMPP in a continuous-space scenario, where agents perform online collision avoidance using ORCA~\cite{vandenberg_springer11}, one of the representative collision-avoidance planners.

\paragraph*{Setup}
The map at the top of Fig.~\ref{fig:ORCA_experiment} shows the environment.
Starts $\mathcal{S}^*=(s_1^*,\dots,s_n^*)$ and goals $\mathcal{G}^*=(g_1^*,\dots,g_n^*)$ are randomly sampled.
A sparse graph $G = (V, E)$ is obtained by placing vertices on a $4\;\text{m} \times 3.5\;\text{m}$ lattice and connecting 4-neighboring vertices.
Each $s_i^*$ and $g_i^*$ is mapped to its nearest vertex, yielding $\mathcal{S},\mathcal{G}$ on $G$.
A-CMTS ($\omega = 1.3$) is then run for \SI{10}{\second} on $(G,\mathcal{S},\mathcal{G})$ to generate CMPP paths~$\pi$.

\paragraph*{Integration of CMPP with ORCA}
For each agent~$i$, we form a waypoint queue $\mathcal{Q}_i=(\pi_i[2],\ldots,\pi_i[L_i-1], g_i^*)$, i.e., the CMPP path without its first and last element, and with the exact goal appended.
ORCA steers the agent toward the current head waypoint $\mathcal{Q}_i^{\text{front}}$.  
When the agent is sufficiently close to $\mathcal{Q}_i^{\text{front}}$ (within a threshold $r$), the front waypoint is popped from $\mathcal{Q}_i$ and the next becomes active. 
We set $r=5.0$ m in this experiment.

\paragraph*{Results}
The bottom row of Fig.~\ref{fig:ORCA_experiment} plots the \textit{Success Rate}, which represents the percentage of agents that have reached their goals at each time step.
With $|A|=200$, both vanilla ORCA and CMPP-guided ORCA reach 100 \% success within \SI{40}{\second}.
At $|A|=300$, agent congestion emerges under vanilla ORCA, delaying convergence to nearly \SI{60}{\second}, whereas CMPP guidance maintains a 100\% success rate after \SI{40}{\second}.
For $|A|=400$, the mean success rate at \SI{60}{\second} improves from 83.9 \% with vanilla ORCA to 99.0 \% with CMPP guidance, a gain of 15.1 percentage points.

\subsection{Multi-Agent Guidance Simulation in Discrete Space} \label{sec:app_lifelong_mapf}
We validate the effectiveness of CMPP in a discrete-space lifelong-MAPF scenario, where each agent receives a new destination immediately upon reaching its current goal.
Local collision avoidance is handled online with PIBT~\cite{okumura2022priority}.
This experimental setup closely resembles the robotics competition \textit{The League of Robot Runners
}~\cite{chan2024the} sponsored by Amazon Robotics, where PIBT serves as the default planner.
The evaluation is conducted in the \textcolor{black}{seven} environments in Fig.~\ref{fig:applied_experiment_results}.
The first two are from~\cite{Ho2024RAL}, while the remaining \textcolor{black}{five} are from the MAPF benchmark~\cite{Stern_SoCS2019}.

\paragraph*{Sparse-Graph Abstraction for CMPP}
A sparse graph $G=(V, E)$ (middle row of Fig.~\ref{fig:applied_experiment_results}) is constructed from the original 4-connected grid $G^*=(V^*, E^*)$ by sampling vertices at fixed intervals.
A surjective map $f:V^*\to V$ assigns each grid cell to its nearest sparse vertex, while an injective map $g:V\to V^*$ links each sparse vertex back to a representative grid cell.

\paragraph*{Integration of CMPP with PIBT}
A-CMTS computes CMPP solutions $\pi_i = (\pi_i[1], \dots, \pi_i[L_i])$ for each agent~$i$ on graph $G$.
Each agent's next waypoint on the grid $G^*$ is determined from this path: if $L_i \geq 3$, the agent is directed to the representative cell of $\pi_i[2]$ via the injective map $g: V \to V^*$; otherwise, it proceeds directly to its goal $g_i$.
PIBT then plans one-step, collision-free moves toward these waypoints.
Since agents advance one grid cell at each time step, their positions on the sparse graph are updated before the next CMPP iteration using the surjective map $f: V^* \to V$.
The visited vertex is then removed from each agent's CMPP path, and the remaining path is reused as the initial solution for the subsequent A-CMTS computation.

\paragraph*{Setup}
Agents are placed randomly on traversable cells and assigned random goals.
Each simulation runs for $500$ steps and is repeated $25$ times under different random scenarios.
A-CMTS with $\omega=1.3$ solves CMPP at each step within a \SI{1}{\second} runtime limit.
We compare the following baseline strategies against CMPP-guided PIBT:
\emph{(i)} \textbf{Vanilla PIBT:} each agent moves directly toward its destination; 
\emph{(ii)} \textbf{Parity-Heuristic:} a simple guide heuristic for PIBT, in which an agent prefers moving up (down) when its current $x$-coordinate is odd (even), and right (left) when its $y$-coordinate is odd (even), thereby promoting left-hand-rule behavior; 
\emph{(iii)} \textbf{TFO:} a state-of-the-art lifelong-MAPF solver (\textit{GP-R100} variant)~\cite{Chen_AAAI2024}.

\paragraph*{Results}
The bottom row of Fig.~\ref{fig:applied_experiment_results} presents the results.
Parity-Heuristic improves vanilla PIBT but offers only marginal throughput gains.
TFO optimizes guidance paths at the grid level, achieving strong performance on warehouse maps, but reducing throughput on \textit{lak303d} and \textit{random-64-64-20}.
CMPP consistently improves PIBT throughput across both warehouse and random maps, raising throughput by $58.1\%$ for \textit{warehouse-10-20-10-2-1} ($|A|{=}1,500$) and by $15.7\%$ for \textit{random-64-64-10} ($|A|{=}1,900$).
On \textit{lak303d}, unavoidable single-lane bottlenecks prevent rerouting; thus, the longer detours imposed by TFO and CMPP slightly reduce throughput compared to vanilla PIBT.
These results confirm that CMPP boosts guidance efficiency in dense environments whenever alternative routes are available.

{
\begin{figure}[t]
    \centering
    \includegraphics[width=\linewidth]{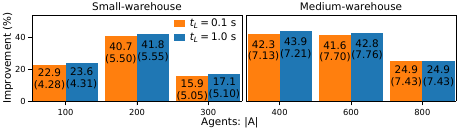}
    \caption{
        Throughput improvements of CMPP-guided PIBT relative to vanilla PIBT for varying A-CMTS runtime $t_L$ (\si{\second}). Bars: improvement (\%); raw throughput in parentheses.
    }
    \label{fig:A_CMTS-runtime}
\end{figure}
}
\paragraph*{Impact of Runtime Limit}
Figure~\ref{fig:A_CMTS-runtime} shows the results of varying the runtime limit, denoted as $t_L$, for A-CMTS.
To observe the effect of $t_L$ clearly, we selected test patterns where A-CMTS can still perform a meaningful number of search expansions under $t_L=0.1$.
Increasing $t_L$ to $1.0$ allows A-CMTS to explore more nodes, improve the solution quality, and achieve higher throughput.


\section{Conclusion} \label{sec:conclusion}
In this paper, we introduced a novel path-planning problem, CMPP, which mitigates congestion for distributed autonomous agents by embedding a flow-based multiplicative congestion penalty directly into the cost function.  
To solve CMPP, we proposed an MINLP-based approach for small instances and A-CMTS for large-scale problems.  
Experiments showed that while the MINLP solver quickly found exact solutions on small instances, it failed to produce feasible ones for large instances.  
In contrast, A-CMTS produced scalable suboptimal solutions within limited computation time.  
Applied experiments confirmed that CMPP reduces agent congestion and improves overall navigation efficiency in both continuous- and discrete-space scenarios.

CMPP is applicable to a wide range of autonomous systems that follow coarse-level route guidance while handling local collision avoidance in a decentralized manner, including traffic management for aircraft and UAVs.
In practical deployments, challenges may arise in scenarios involving agents with unpredictable behaviors; thus, ensuring robustness in such environments remains an open issue.
Future work includes integrating CMPP with related optimization problems, such as product placement in logistics warehouses and dynamic task allocation.


\bibliographystyle{IEEEtran}
\bibliography{ref_avrb}

\end{document}